\documentclass[fleqn,usenatbib]{mnras}

\usepackage{newtxtext,newtxmath}

\usepackage[T1]{fontenc}
\usepackage{ae,aecompl}

\usepackage{graphicx}	
\usepackage{amsmath}	
\usepackage[normalem]{ulem}
\usepackage{multirow}
\usepackage{caption}
\usepackage{xcolor}

\title[Photometric age and metallicity from CNNs]{Constraining stellar population parameters from narrow band photometric surveys using convolutional neural networks}

\author[C. L. Liew-Cain et al.]{
Choong Ling Liew-Cain,$^{1}$\thanks{E-mail: choongling.liew-cain.18@ucl.ac.uk (CLLC)}
Daisuke Kawata,$^{1}$
Patricia S\'{a}nchez-Bl\'{a}zquez,$^{2}$
\newauthor{Ignacio Ferreras,$^{1, 3, 4}$}
Myrto Symeonidis$^1$
\\
$^{1}$Mullard Space Science Laboratory, University College London, Holmbury St Mary, Dorking, RH5 6NT, UK\\
$^{2}$Departamento de F\'{\i}sica de la Tierra y Astrof\'{\i}sica, Universidad Complutense de Madrid, 28040, Madrid\\
$^{3}$Instituto de Astrof\'{i}sica de Canarias, Calle V\'{i}a L\'{a}ctea s/n, E38205, La Laguna, Tenerife, Spain\\
$^{4}$Departamento de Astrof\'{i}sica, Universidad de La Laguna (ULL), E-38206 La Laguna, Tenerife, Spain
}

\date{Accepted XXX. Received YYY; in original form ZZZ}

\pubyear{2019}

\begin{document}
\label{firstpage}
\pagerange{\pageref{firstpage}--\pageref{lastpage}}
\maketitle

\begin{abstract}
Upcoming large-area narrow band photometric surveys, such as J-PAS, will enable us to observe a large number of galaxies simultaneously and efficiently. However, it will be challenging to analyse the spatially-resolved stellar populations of galaxies from such big data to investigate galaxy formation and evolutionary history.
We have applied a convolutional neural network (CNN) technique, which is known to be computationally inexpensive once it is trained, to retrieve the metallicity and age from J-PAS-like narrow band images. 
The CNN was trained using mock J-PAS data created from the CALIFA IFU survey and the age and metallicity at each data point, which are derived using full spectral fitting to the CALIFA spectra. 
We demonstrate that our CNN model can consistently recover age and metallicity from each J-PAS-like spectral energy distribution. The radial gradients of the age and metallicity for galaxies are also recovered accurately, irrespective of their morphology. However, it is demonstrated that the diversity of the dataset used to train the neural networks has a dramatic effect on the recovery of galactic stellar population parameters. Hence, future applications of CNNs to constrain stellar populations will rely on the availability of quality spectroscopic data from samples covering a wide range of population parameters.
\end{abstract}




\section{Introduction}

The determination of the stellar population properties in galaxies is one of the most powerful techniques to understand the formation and evolution of galaxies. Traditionally, this has been done by comparing the line spectral features with stellar population synthesis models \citep[e.g.][]{Worthy94, Bruzual03, Vazdekis10, Conroy13}, using spectral indices \citep[e.g.][]{Trager00, SanchezBlazquez16} or, more recently, using full spectral fitting techniques \citep[][]{Panter03}.

Over the past few years, galactic spectra have been obtained by Integral Field Unit (IFU) surveys, including Calar Alto Legacy Integral Field Area \citep[CALIFA,][]{CALIFA}, Mapping Nearby Galaxies at APO \citep[MaNGA,][]{MaGNA}, Sydney-Australian-Astronomical-Observatory Multi-object Integral-Field spectrograph \citep[SAMI,][]{SAMI}, \textit{K}-band Multi Object Spectrograph \citep[KMOS,][]{KMOS3D}. These IFU surveys can be used to produce two-dimensional distributions of age and metallicity to be studied for different galaxy types. 
These spatially resolved spectra have put strong constraints on galaxy formation and stellar population synthesis models \citep[e.g.][]{Belfiore19}.

An alternative to spectroscopic surveys comes from narrow band filter imaging. Photometric surveys are more efficient at observing fainter objects than spectroscopic instruments, and can cover a greater area on the sky in a single observation.
In  photometric surveys, galaxies are not pre-selected, unlike in spectroscopic surveys. Instead, all galaxies that are brighter than the limiting magnitude in the field of view will be observed. Narrow and medium band filter surveys, such as Classifying Objects by Medium-Band Observations \citep[COMBO-17,][]{COMBO17}, Survey for High-z Absorption Red and Dead Sources \citep[SHARDS,][]{SHARDS}, Javalambre Physics of the Accelerating Universe Astrophysical Survey \citep[J-PAS,][]{JPAS}, Javalambre Photometric Local Universe Survey \citep[J-PLUS,][]{JPLUS} and Southern Photometric Local Universe Survey \citep[S-PLUS,][]{SPLUS}, effectively act as low spectral resolution IFU surveys, producing spectral energy distributions (SEDs) at many positions within the galaxy. These SEDs contain enough information to derive an average stellar age and metallicity 
\citep[e.g.][]{SanRoman18}.
For example, \citet{DiazGarcia15} used Advanced Large Homogeneous Area Medium Band Redshift Astronomical Survey (ALHAMBRA) data to derive redshift, metallicity and age and compare these values with spectroscopic observations of the same galaxies. The Multi-Filter Fitting for stellar population diagnostics \citep[MUFFIT, ][]{DiazGarcia15} code they developed shows good recovery of the spectroscopic values, though results are highly dependent on the choice of stellar population model.
\citet{SanRoman19} analyses two elliptical galaxies observed by J-PLUS. The radial gradients for age, metallicity and extinction that are derived are in reasonable agreement with CALIFA survey observations of the same galaxies.

A challenge emerging from narrow-band surveys is the volume of data to be analysed. For example, J-PAS aims to observe a total of $9\times 10^7$ galaxies with multiple pixels per galaxy. Additionally, J-PAS and J-PLUS together are expected to collect a maximum of 1.5 TB of data per night \citep{JPAS}. Therefore, a computationally efficient method for deriving stellar population parameters from the data is required, and will become invaluable in the future with larger surveys. In this paper, we present neural networks as a tool that shows promise in overcoming this challenge.

Neural networks are algorithms that allow non-linear mapping between input and target parameters, and are efficient methods of analysing large datasets. 
Supervised machine learning uses an input dataset, such as  photometric SEDs, and the set of "true" values of the target parameter, e.g. age or metallicity, to learn how to make accurate predictions. Selecting an appropriate training set is a vital step in neural network  methods. Galaxies have a diverse formation history and therefore the training set needs to cover this wide variety of galaxy evolution. Otherwise, the neural network will not be capable of accounting for the diversity present in galactic surveys.

Machine learning is applied widely in astrophysical research \citep[e.g.][]{Folkes96, AstroML} and has been used to derive the metallicity of galaxies from broad band photometric surveys previously. \citet{Acquaviva16} and \citet{WuBoada} applied random forest algorithms and neural networks respectively to calculate the metallicity of galaxies from multi-wavelength Sloan Digital Sky Survey (SDSS) photometric observations, with SDSS spectral age and metallicities used as training data. \citet{Lovell19} used the results of cosmological simulations of galaxies to synthesise SDSS-like spectra. The authors included simulated effects of extinction and noise when creating these SEDs. Convolutional neural networks (CNNs) were trained on these SDSS-like spectra to determine galactic star formation rate over cosmic time. 
\citet{WuBoada} noted that increasing the number of photometric filter bands used to train the neural network improved the accuracy of the predicted metallicity value of the galaxy. Therefore, the application of neural networks to narrow band photometric surveys, as in this paper, is an obvious step in deriving galactic evolution parameters.
This paper is a proof of concept study, investigating whether neural networks can be used to derive the age and metallicity parameters from narrow-band photometric data. We also examine how the accuracy of recovering age and metallicity gradients, compared to those derived directly from the spectra, depends on the training set use in the neural network.

In the next section, the synthesis of the data is discussed. This is followed by the methodology of the neural network and analysing gradient retrieval in Section \ref{sec:method}. Section \ref{sec:results} presents the results of the neural network. Discussion and conclusions are provided in Section \ref{sec:conclusion}.

\section{Data}
\label{sec:data}

In this paper, we develop a neural network model to derive metallicity from the narrow-band filter photometric data, similar to the data which will potentially be gathered by the J-PAS survey. We targeted the J-PAS survey because it is the next generation large scale survey, and a computationally efficient analysis tool is required to derive stellar population properties for the many pixels of  photometric data. To this end, we construct J-PAS-like narrow band filter data, i.e. 'mock J-PAS data', from CALIFA IFU spectra. We then assume that the spectroscopically derived ages and metallicities from the CALIFA data are the true values for each spectrum within each galaxy. The training and testing datasets for our neural network are composed of the mock J-PAS data and the spectroscopically derived age and metallicity. In Section \ref{subsec:CALIFA} we explain the CALIFA data, and in Section \ref{subsec:synthesis} we describe how we make the synthesised J-PAS data from the CALIFA spectra.

\subsection{CALIFA}
\label{subsec:CALIFA}
The CALIFA survey \citep{CALIFA} used the PMAS/ PPAK integral field spectrograph, mounted on the Calar Alto 3.5 m telescope.
Each galaxy in the dataset was observed three times, with dithering used to reach a spectral resolution of $\sim 1$". 
The integral field unit (IFU) allows 2D spectra in a grid over the surface of the galaxy to be collected, through exposure times of 1800 s and 900 s for the blue and red gratings respectively.
The CALIFA parent sample consists of 937 galaxies selected from SDSS DR7 within $0.005 < z < 0.03$, with the majority being field galaxies. 
From the parent sample, $\sim 600$ galaxies were observed with a diameter limit to fit within the IFU field of view and down to $M_B \sim -18.0$ mag
\citep[for more information about the CALIFA sample see][]{CALIFA, Walcher14}.

Star formation histories were derived using the code STEllar Content and Kinematics via Maximum A Posteriori likelihood \citep[STECKMAP,][]{Ocvirk06a} on the emission line-cleaned spectra using  \citet{Vazdekis10} stellar population models, with the MILES stellar library \citep{SanchezBlazquez06b}, a Kroupa Universal IMF \citep{Kroupa01} and Padova 2000 \citep{Girardi00} isochrones, which cover a range of ages and metallicities from 
63 Myr to 17.8 Gyr and 
$ - 2.32 < $[Z/H]$ < +0.2$ respectively 
\citep[for a detailed description of the procedure see][]{SanchezBlazquez14}. No cosmological priors were applied when the values for the ages of the stellar populations were determined. This means that the ages of the galaxies are allowed to be, in principle, higher than the age of the Universe.

We have decided to use IFU data as it is the most suitable for radial gradient analysis of galaxies. IFU data allows better spatial averaging of galactic properties than long slit instruments. The sample used in this analysis comprises a total of 190 galaxies with high enough quality data to compute age and metallicity. Of this sample, 44 galaxies are early-type galaxies and 146 are late-types according to their classification on the SIMBAD database \citep{SIMBAD}. This is not representative of the full CALIFA sample \citep{Walcher14} which contains a significantly higher fraction of elliptical galaxies. From the star formation history and age -- metallicity relation derived with STECKMAP, we calculate a mean luminosity weighted age and metallicity for each spectrum in the dataset using spectral fitting. Any spectra whose fit was deemed to be poor (i.e. with reduced $\chi^2 > 2$) were ignored for this work, 
giving a dataset composed of 19,727 spectra. 



\subsection{Synthesised J-PAS data}
\label{subsec:synthesis}


The J-PAS survey is a multiband photometric survey which will run at the Observatoro Astrofisico de Javalambre in Spain, with a $3.89 m^2$ collecting mirror. The J-PAS instrument covers a 4.7 square degrees per observation, with a pixel size of 0.456 arcsec. The effective integration time is 4.96 hours per field \citep{JPAS}. 

The response curve of the 54 narrow-band filters are spaced 100 {\AA} apart with a FWHM of 145 {\AA}, covering the range of $3785-9100$ {\AA}. The magnitude limit is $21.0 < m_{AB} < 25.7$ mag, and varies by filters. These narrow band filters act as a low-resolution spectrograph, with an effective resolution of 100 {\AA} (compared to CALIFA's resolution of 2{\AA}) and are able to detect the broad galaxy emission features. 

\begin{figure}
\includegraphics[width=\columnwidth]{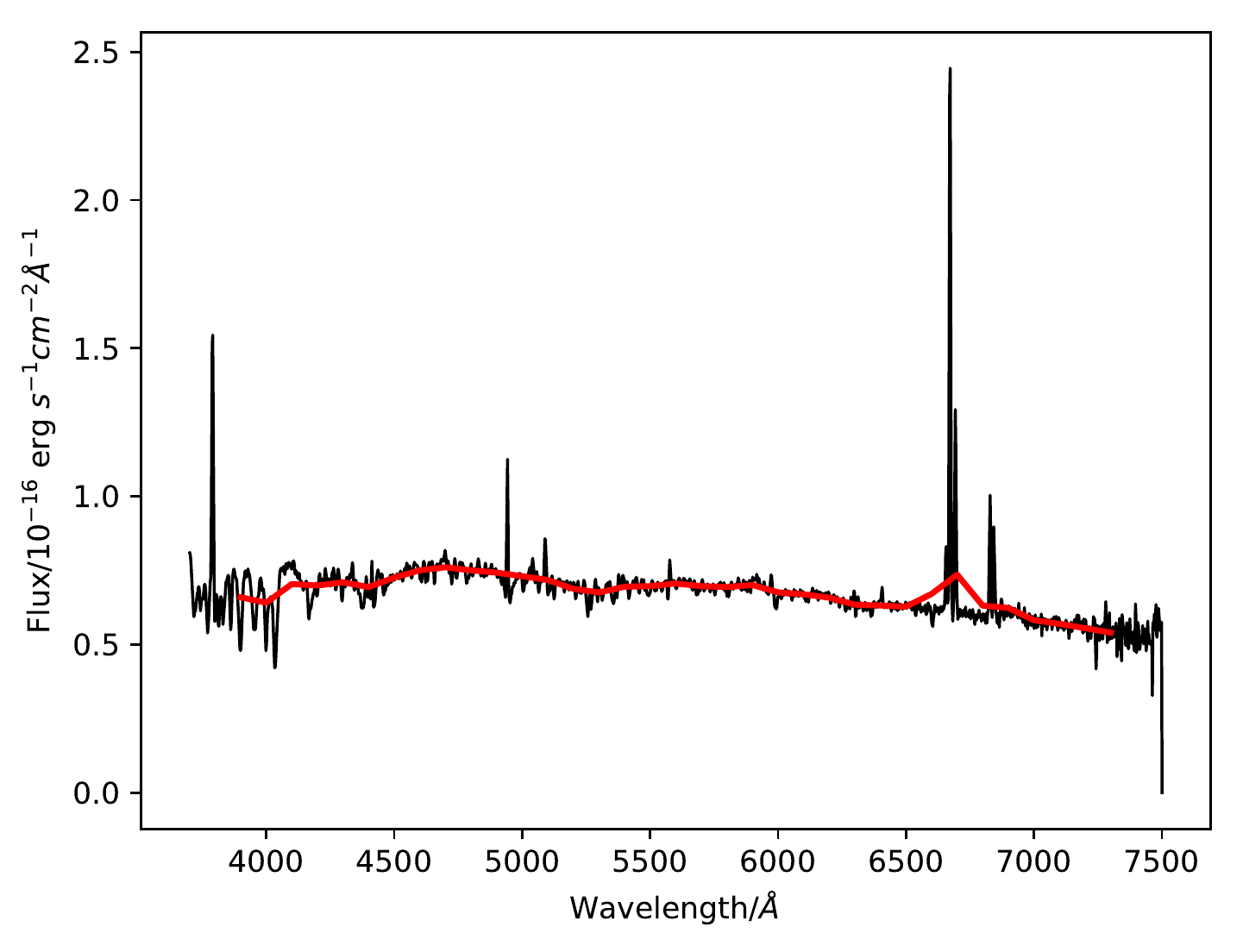}
    \caption{A comparison of the spectral curve given by CALIFA (black) and the simulated J-PAS response (red) for one spectrum in NGC2530. The majority of spectral lines cannot be seen in the J-PAS SED, making it more difficult to extract age and metallicity information. }
    \label{fig:SED}
\end{figure}

The mock SEDs are created by convolving the CALIFA spectral data from each point with the known response functions of the J-PAS filters. As the spectral range of CALIFA is 3700-7000 {\AA} only 36 J-PAS-like bands are  constructed from the intersection of the two instruments' wavelength ranges for use in this analysis. An example of the generated mock J-PAS SED and the original CALIFA spectrum can be seen in Fig. \ref{fig:SED}, where the red line shows the mock J-PAS SED.  The black curve shows the full, cleaned CALIFA spectrum. The lack of absorption line features in the narrow band SED has previously made the determination of age and metallicity significantly more challenging for photometric instruments compared to spectral surveys. We masked the H$\alpha$ line at $\lambda \sim 6563$ {\AA} 
as contamination from nebular emission complicates the analysis of stellar populations.

\section{Method}
\label{sec:method}

\subsection{Neural network}
\label{subsec:NN}

We use supervised neural networks to predict the metallicity and age of a sample of galaxies from their J-PAS-like SEDs (see Section \ref{subsec:synthesis} for details on their synthesis) with the Tensorflow \textit{Keras} API \citep{Tensorflow}\footnote{See https://github.com/ChoongLing/SimulatedJ-PAS for the code used for the methods discussed in this section.}. The convolutional neural network (CNN) we develop uses the spectroscopic age and metallicity derived by CALIFA as the 'true' value for the purposes of training. Each of the neurons in the network begins with some randomized weight, and the simulated magnitudes for each band pass through the CNN to calculate a predicted value for the age or metallicity. The mean squared error of predicted versus spectroscopic age or metallicity is back propagated through the network to adjust the weights of the neurons. This process is repeated to obtain an accurate output. 

\begin{figure}
\includegraphics[width=\columnwidth]{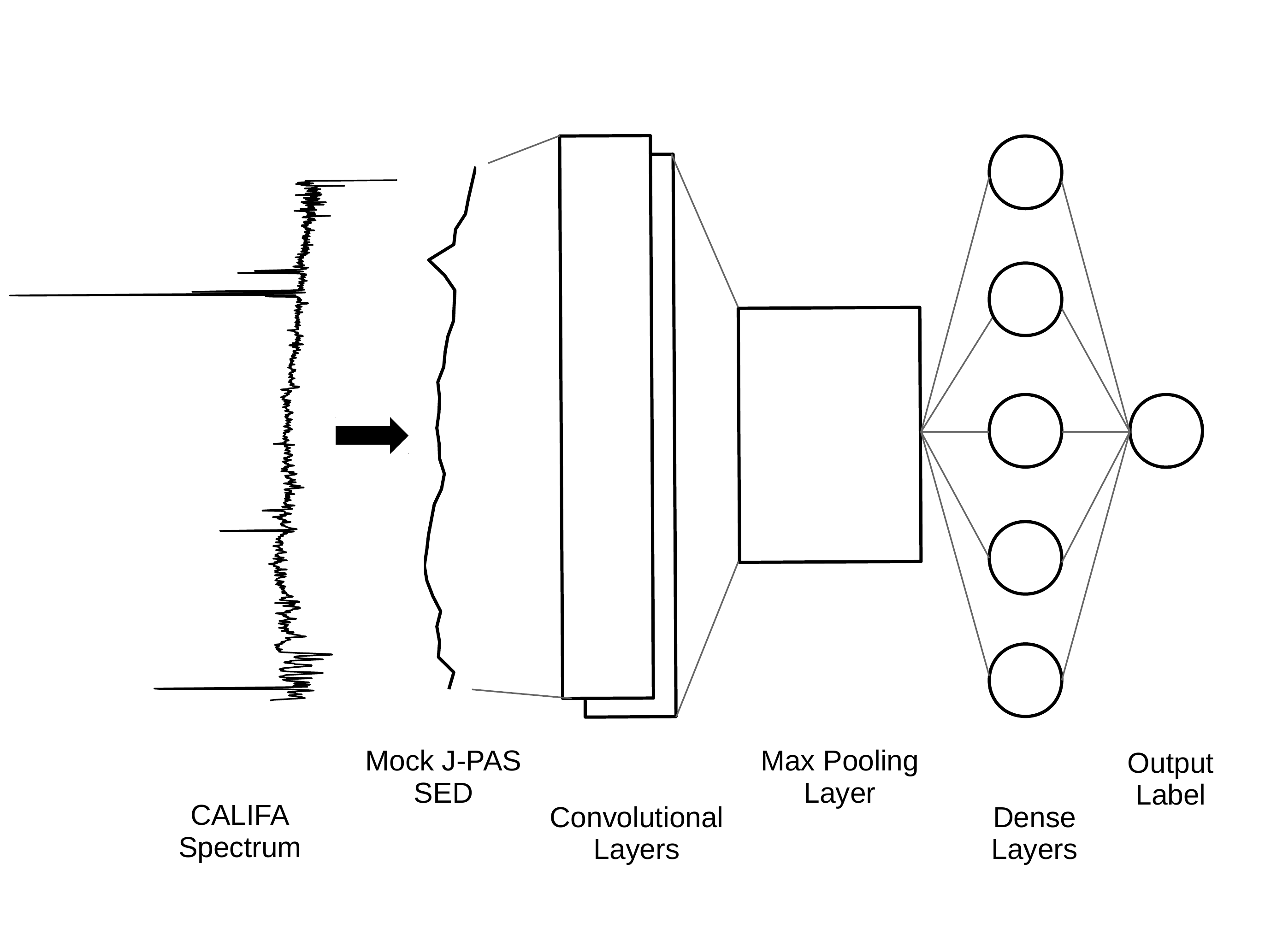}
    \caption{A schematic view of the architecture used for the convolutional neural network (CNN). The CALIFA spectra are converted into mock J-PAS photo-SED, which are then passed through two convolutional layers. A max pooling layer reduces dimensionality, and its results are passed through a single dense layer. The predicted value of age or metallicity is then output by the CNN.}
    \label{fig:NNLayout}
\end{figure}

The CNN used in this work has an architecture as illustrated in Fig. \ref{fig:NNLayout}. The starting point for the CNN was taken from \citet{Fabbro18}, who used a CNN to analyse stellar spectra. Our chosen architecture has two convolutional layers, a max pooling layer and a single dense layer. The 1D convolutional layers capture patterns and multi-filter features across the SED. The max pooling layer then reduces the dimensions of the convolutional layers' output. This is applied to the classical dense neural network layer which calculates the age or metallicity via non-linear combinations of values given by the outputs of the max pooling layer.
The age and metallicity were determined by separate CNN models, which had identical architectures but different hyperparameters, which are shown in Table \ref{tab:hyperparameters}. The layers' hyperparameters were optimised by Hyperas\footnote{https://github.com/maxpumperla/hyperas}. Comparisons showed that the set of hyperparameters chosen by Hyperas provide more accurate predictions than are made by CNNs with manually chosen hyperparameters. 

\begin{table}
  \centering
	\caption{The hyperparameters used in the CNN. (1) and (2) indicate the first and second convolutional layer respectively.}
	\label{tab:hyperparameters}
	\begin{tabular}{lcc} 
		\hline
		\textbf{Layer Parameter} & \textbf{Age} & \textbf{Metallicity}\\
		\hline
		\textbf{1D Conv} & & \\
		Filters (1) & 16 & 33 \\
		Kernel size (1) & 6 & 12\\
		Filters (2) & 16 & 50\\
		Kernel size (2) & 8 & 10\\
		\textbf{Max Pooling} &  & \\
		Pool size & 8 & 2\\
		\textbf{Dense} & & \\
		Neurons & 40 & 33 \\
		\hline
	\end{tabular}
\end{table}

We also adopted early stopping with a patience parameter of 250 for the CNN. This meant that if there was no improvement in the mean absolute error of the parameter recovery after 250 epochs, training would stop. The CNN would train for a maximum of 5000 epochs or until the error stabilised. A total of 19,727 spectra from 190 of galaxies was used in this analysis.

To train the neural network to predict metallicity and age for the full dataset, 25\% of the data was kept aside for the testing of the trained CNN to produce our results. The other three quarters was used for training the CNN.  This process was repeated three more times so that metallicity and age predictions were made for the full dataset, with each iteration using a training set independent of the unseen testing set.

\subsection{Defining the Training and Testing Sets}
\label{subsec:definingsets}

\begin{figure}
\centering
\includegraphics[width=\columnwidth]{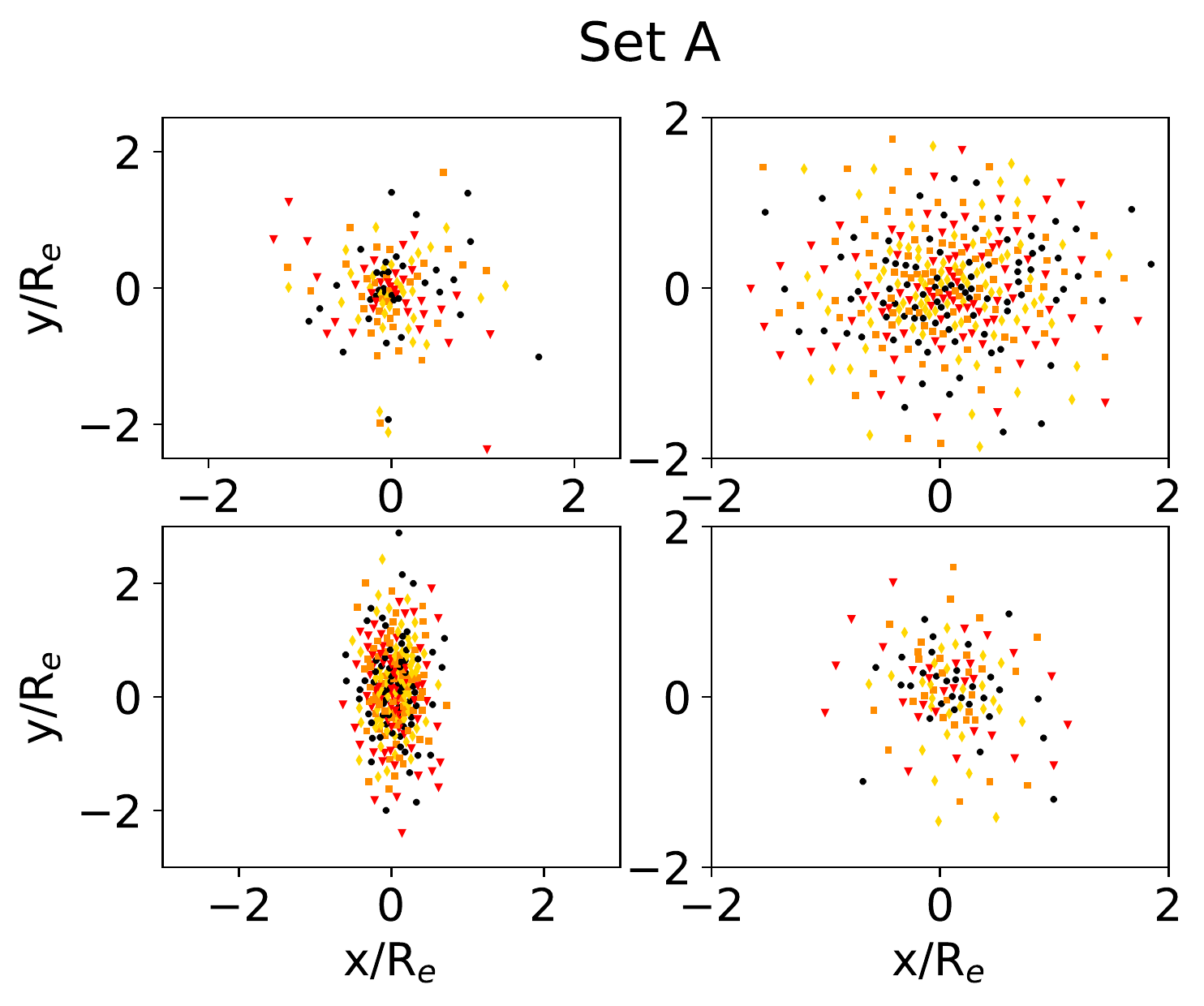}
\includegraphics[width =\columnwidth]{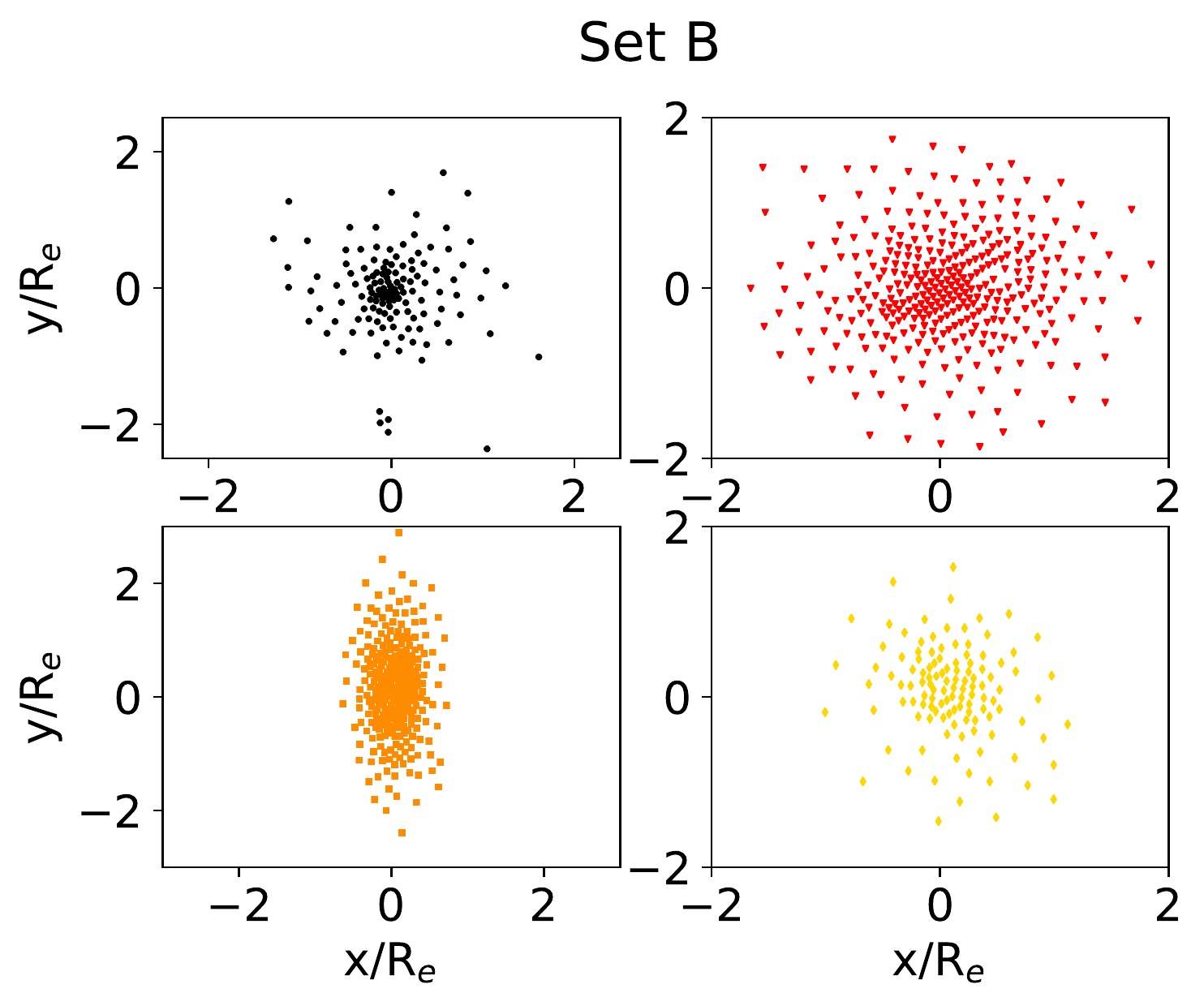}
    \caption{Illustrations showing how the spectral data are split into four subsets, as described in Section \ref{subsec:definingsets}. The top four panels show the splitting for Set A and the lower four for Set B. In both sets of panels, the spatial distribution of the spectra in four different galaxies are shown. Each spectrum is represented by a coloured shape depending on which subset it belongs to (black circles, red triangles, orange squares or yellow diamonds). In Set A, the spectra within each galaxy are split amongst the four subsets, whereas in Set B all of the spectra for a given galaxy are in the same subset.}
    \label{fig:SetAB}
\end{figure}

Two ways of splitting the dataset into four subsets are explored in this work, which are illustrated in Fig. \ref{fig:SetAB}. The first is by splitting the spectra within each galaxy randomly into the four subsets, ensuring that one quarter of the data from each galaxy are put into each one of the four subsets. The CNN is then trained on three of the four subsets, with the final subset kept aside and unseen for testing. This will be referred to as Set A. The other method, Set B, is created by randomly splitting the 190 galaxies into four subsets, with all of the spectra from one galaxy in the same subset. This means that the testing set for Set B contains galaxies which have not been seen at all by the CNN during training. The key difference is that in Set A the training set contains spectral data from every galaxy, therefore the training and testing datasets are not completely independent due to the covariance between adjacent spectra. 

It is possible that spectra from the same galaxy will have similar stellar and chemical evolution histories, even at different positions within the galaxy. In this way, Set A mimics a situation where a large number of galaxies are included in the training set, which will cover the diversity in galactic evolutionary history, so that the training set contains data from similar galaxies to those in the application set. Set B demonstrates the realistic case, where we do not have any previous knowledge about a galaxy in the testing set. In this proof of concept study, we compare the ideal case of Set A with the realistic case in Set B. Although it is more realistic, Set B suffers due to the relatively small size of our dataset. Conversely, Set A is a suitable way of exploring the potential benefits of a large, comprehensive training dataset. Therefore, this comparison will show the potential of the CNN method when a large dataset becomes available in the future.

\subsection{Radial Gradient Analysis}
\label{subsec:derivinggrads}

Radial gradients for the age and metallicity within the effective (half-light) radius, $R_{e}$ of the galaxy are also calculated and analysed for both the CNN predictions (Section \ref{subsec:NN}) and CALIFA spectroscopic age and metallicity. We analysed the gradients only for the galaxies that have at least 25 spectral data points within $R < R_{e}$ and there is at least one data point at $R > R_{e}$, to ensure that enough spectra to cover up to $R < R_{e}$. This allows us to produce reliable radial gradients.

\begin{figure}
\includegraphics[width=\columnwidth]{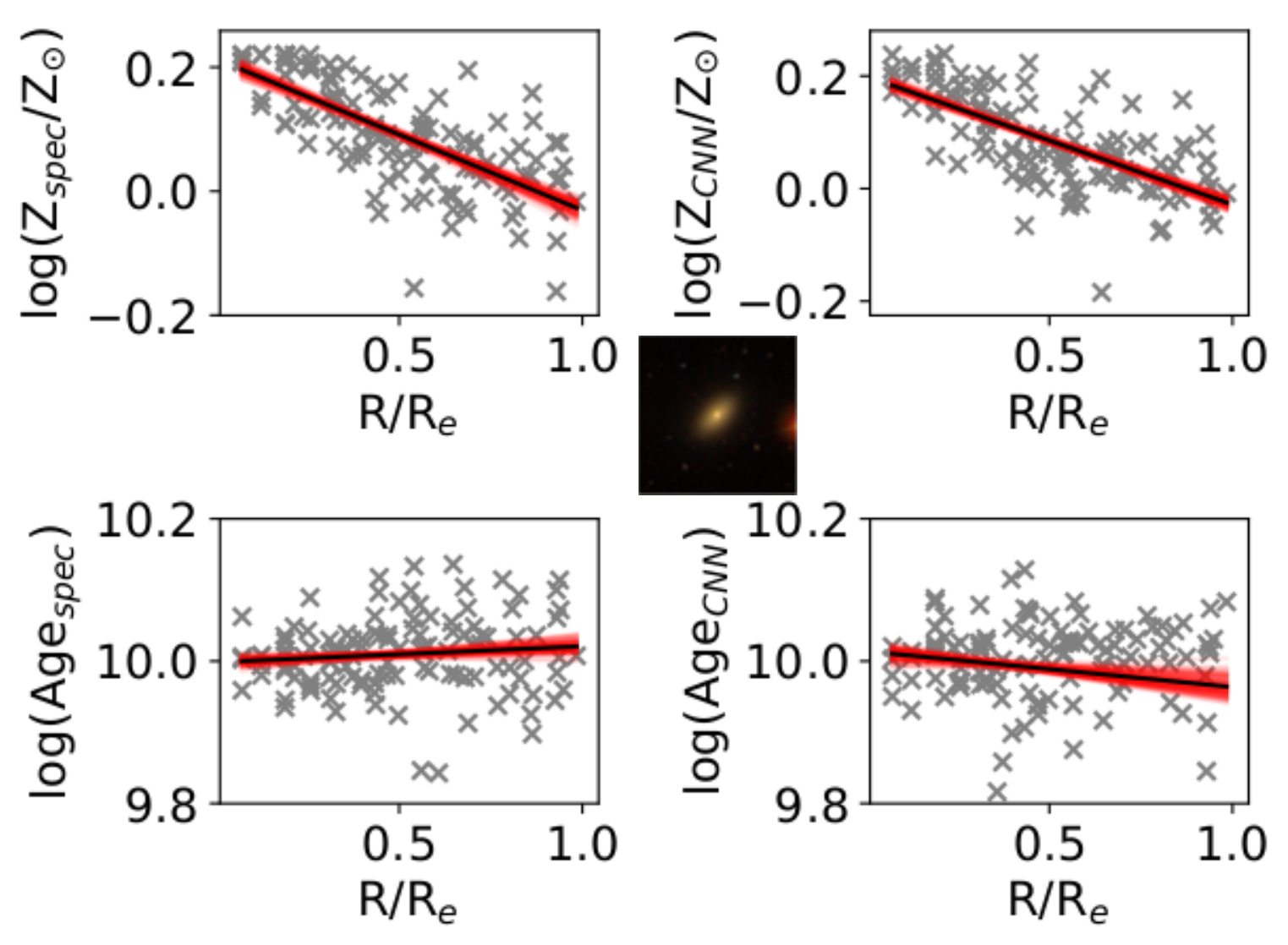}
    \caption{The derived spectroscopic and CNN-predicted ages and metallicities against radius for NGC 7671 with a 4'$\times$4' SDSS image embedded in the centre. The CNN trained using Set A (see Section \ref{subsec:definingsets}). The top row shows metallicity and the bottom panels display age. The left hand column shows the parameter values derived directly from CALIFA spectra, and the right contains predictions from the CNN. The value of each spectrum is shown as grey crosses. The linear fits to these data computed by 100 iterations of MC bootstrapping are shown as red lines, with the mean values for these fits plotted as the solid black line.}
    \label{fig:GradientNGC}
\end{figure}

To obtain the gradient, the inclination of each galaxy was corrected to determine the face-on projected radius for the position of each spectrum. A linear fit to age or metallicity against radius was computed using 
Monte Carlo (MC) bootstrapping to randomly select a sample of 75\% of the data. A least squares fit was obtained for 100 MC samples. Then, the mean gradient and its standard deviation were calculated from these samples. This was performed on both the spectroscopic and CNN predicted values, which were then compared. As no uncertainties were computed from the CNN predictions or spectroscopic values, the uncertainty in the gradient fitting was determined by the standard deviation of the MC derived gradients. Therefore, the uncertainties in the linear gradient fitting do not consider any intrinsic uncertainties in the CALIFA spectroscopic analysis or CNN predictions. 
Fig. \ref{fig:GradientNGC} shows an example where metallicity and age are plotted against radius for the galaxy NGC 7671 using Set A (see Section \ref{subsec:definingsets}). The top row shows the spectroscopic (i.e. the true label, left) and CNN (predictions, right) metallicity, with the bottom row showing the equivalent diagrams for age. The grey crosses are the values for each spectrum. The red lines show the fits produced by each iteration of the MC bootstrapping. The black line shows the gradient derived from the mean value of the MC fits.
The results of gradient analysis will be discussed in Sections \ref{subsec:grads} and \ref{subsec:trainsets}. Only the gradients will be discussed in this paper.

\section{Results}
\label{sec:results}


Results from Set A will be discussed in Sections \ref{subsec:reproducing} and \ref{subsec:grads} and results from Set B will be presented in Section \ref{subsec:trainsets}. We investigate the effects of galactic morphology and training set size on the accuracy of CNN predictions in Section \ref{subsec:morph} and Section \ref{subsec:size}, respectively. Section \ref{subsec:MassVGrads} covers the dependence of our radial gradients on stellar mass.

\subsection{Set A: Predictions of age and metallicity}
\label{subsec:reproducing}

\begin{figure}
    \includegraphics[width=\columnwidth]{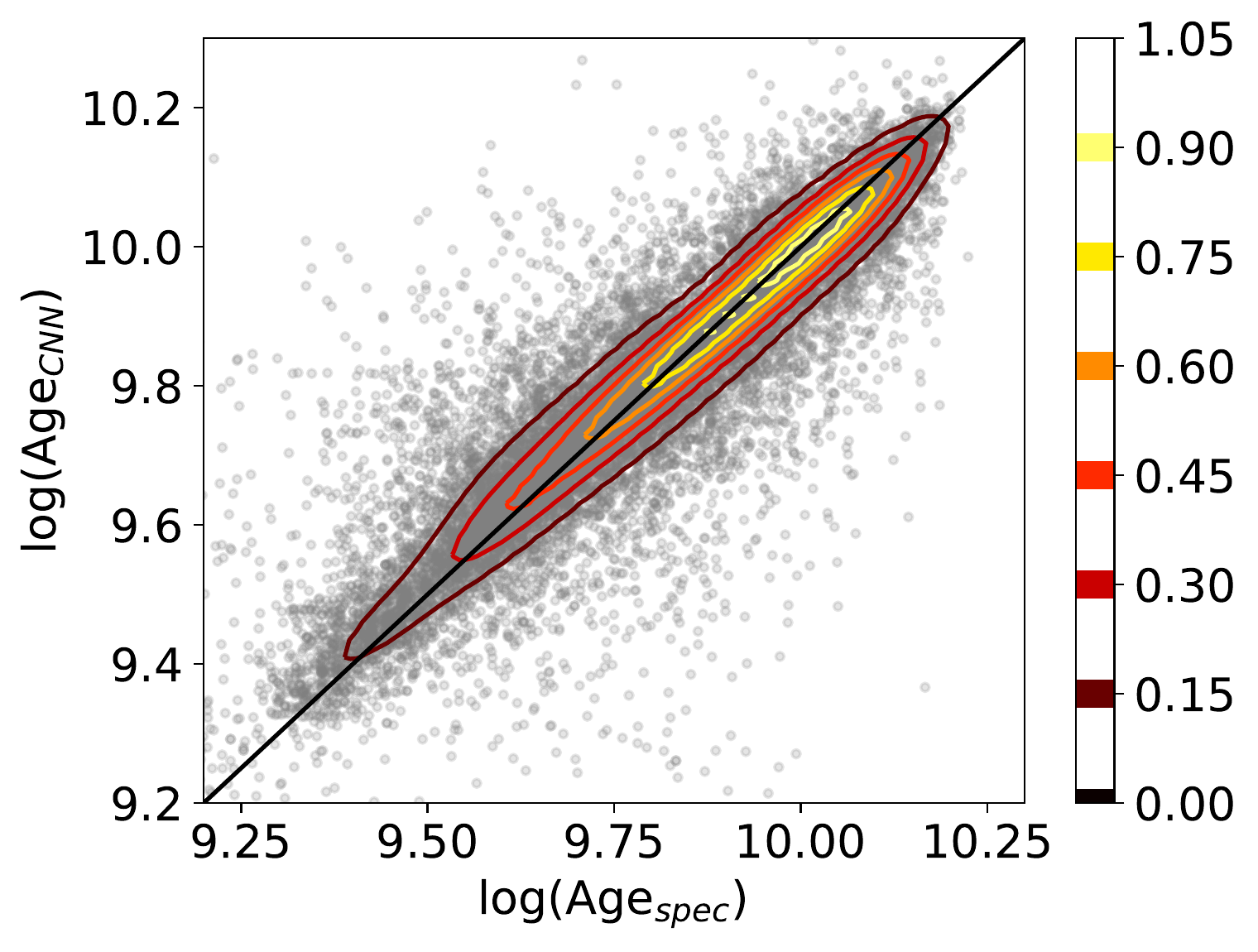}
    \includegraphics[width =\columnwidth]{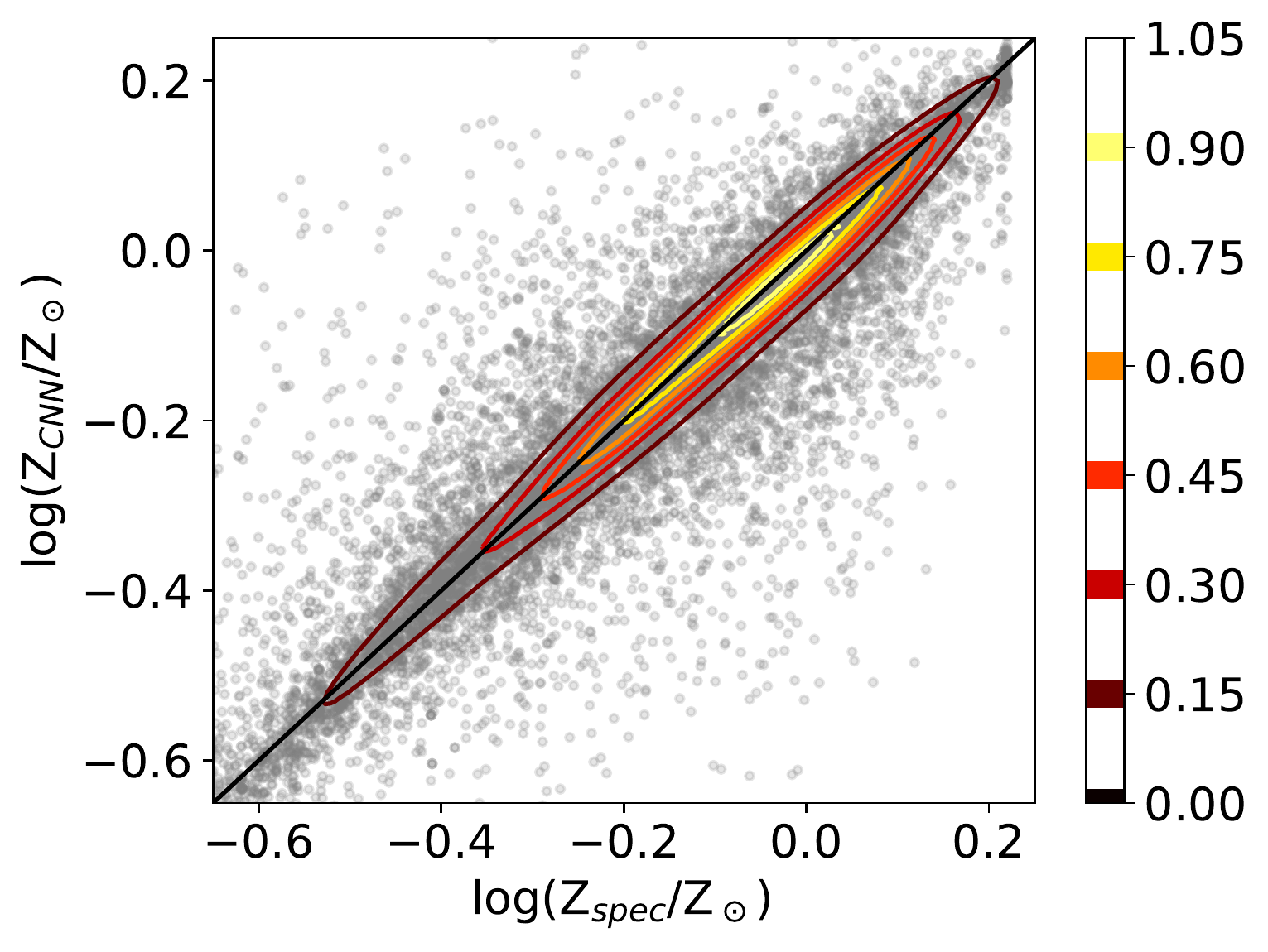}
    \caption{The luminosity-weighted age (top, Age$_{CNN}$) and metallicity (top, Z$_{CNN}$) derived from the CNN against the spectroscopically determined age (Age$_q{spec}$) and metallicity (Z$_{spec}$) for Set A showing only data with a spectroscopically determined value of age and metallicity with reduced $\chi^2 < 2$. The solid black line shows a 1:1 correlation, which corresponds to perfect recovery. The contour map shows the normalised density distributions of the results of the spectra. The CNN values of age and metallicity are consistent with the spectroscopically determined values, with a robust standard deviation of 0.05 and 0.03 dex respectively.}
    \label{fig:ageZrecovery}
\end{figure}

The recovery of age and metallicity using Set A is shown in Fig. \ref{fig:ageZrecovery}. The grey points show the prediction of the CNN against the value determined from CALIFA, which we consider to be the true values. A contour map shows the normalised distribution of these points. The solid black line shows a 1:1 correlation, i.e. a CNN prediction that is identical to the spectroscopic value. The recovery here is excellent, which can be seen as most points lie close to the 1:1 recovery line. The robust standard deviation (calculated from the median absolute deviation) of the difference between CNN and spectroscopic values for Set A are 0.05 dex for age and 0.03 dex for metallicity.

This level of accuracy in reproducing age and metallicity is encouraging, and shows that the CNN is working well. Once the model has been trained, its application to the test dataset is very rapid, meaning it is suitable for use in the large datasets, such as those that will be produced by J-PAS. The standard deviation in the CNN predictions is comparable to those obtained by CALIFA spectral fitting \citep[e.g.][]{SanchezBlazquez14}. 

\subsection{Set A: Gradient analysis}
\label{subsec:grads}

\begin{figure}
    \includegraphics[width=\columnwidth]{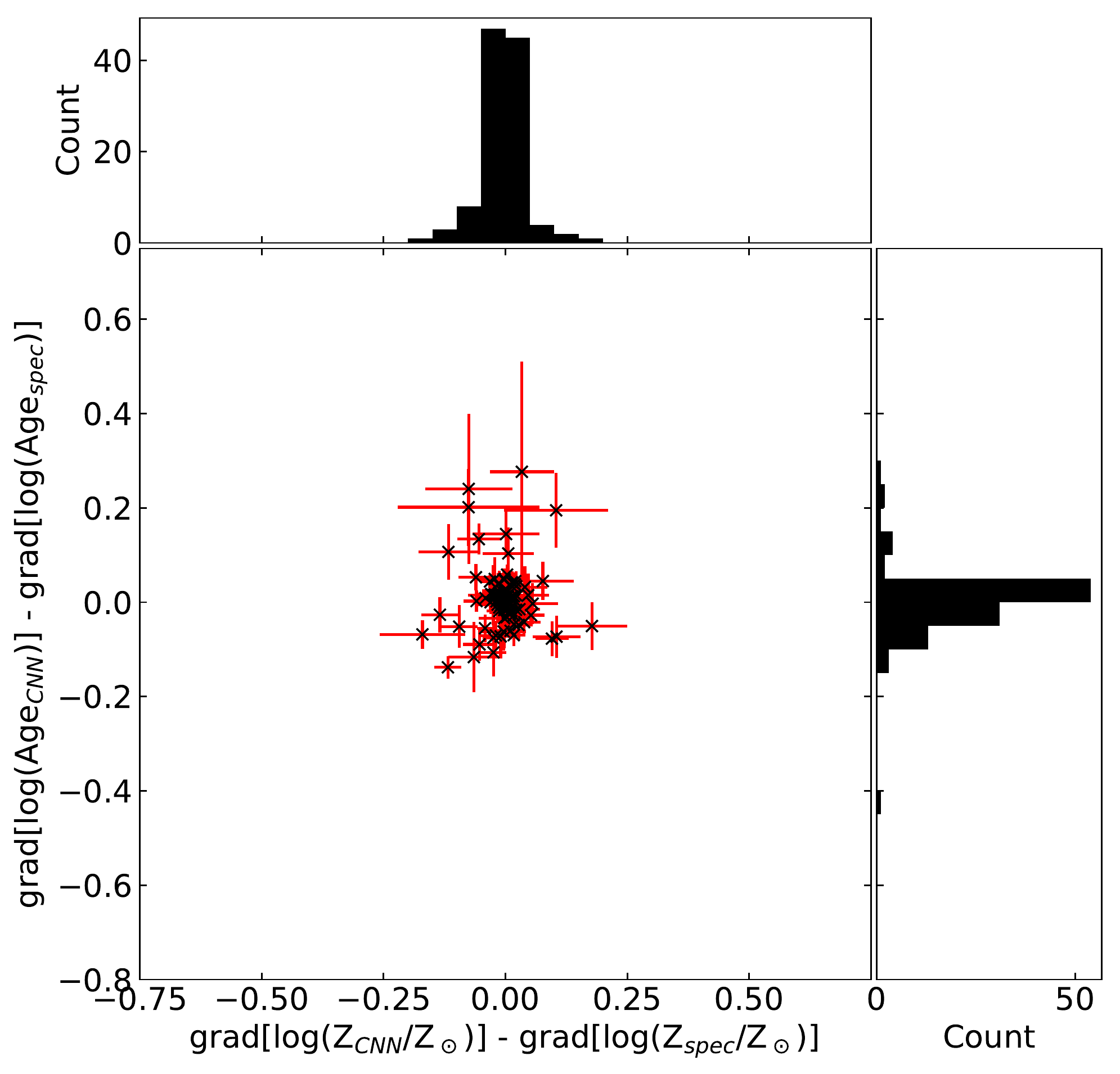}
    \caption{The difference between the gradients from CNN predicted age, grad(log(Age$_{CNN}$)), and the spectroscopically derived age, grad(log(Age$_{spec}$)), against the difference in the CNN predicted metallicity gradient, grad(log(Z$_{CNN}$/Z$_\odot$)), and spectroscopically derived metallicity, grad(log(Z$_{spec}$/Z$_\odot$)). Red error bars show 1-$\sigma$ confidence limits for the gradient fitting. The top and right panels show histograms of the gradient differences in bins of 0.05 dex/$R_{e}$ . The robust standard deviation for the difference in gradients is 0.03 and 0.02 dex/$R_{e}$ for age and metallicity, respectively. There is no visible correlation between differences in CNN predictions for age and metallicity gradient and the respective spectroscopic gradients.}
    \label{fig:ScatteredHists}
\end{figure}

The values of age and metallicity from each point -- both spectral and CNN predicted -- are used to calculate a radial gradient, as described in Section \ref{subsec:derivinggrads}. 
The differences between the CNN predicted and spectroscopic gradients are plotted in Fig. \ref{fig:ScatteredHists}. The black crosses show the difference between the calculated gradients, with the red lines showing 1-$\sigma$ error bars computed using the MC bootstrap sampling. The top and right panels show histograms of the difference between the gradients of metallicity and age, respectively, with bins of 0.05 dex/$R_{e}$. There is strong clustering of the differences in gradient in the central 0.1 dex/$R_{e}$. The gradient recovery is found to be accurate to within a robust standard deviation of 0.03 dex/$R_{e}$ and 0.02 dex/$R_{e}$ in age and metallicity, respectively. It can also be seen that there is no clear correlation between the age and metallicity gradient deviations of the CNN values from the spectroscopic gradients, which shows that the quality of CNN predictions are not affected by the age-metallicity degeneracy.

\subsection{Set B: Age and metallicity prediction and gradient analysis}
\label{subsec:trainsets}

\begin{figure}
    \centering
    \includegraphics[width=\columnwidth]{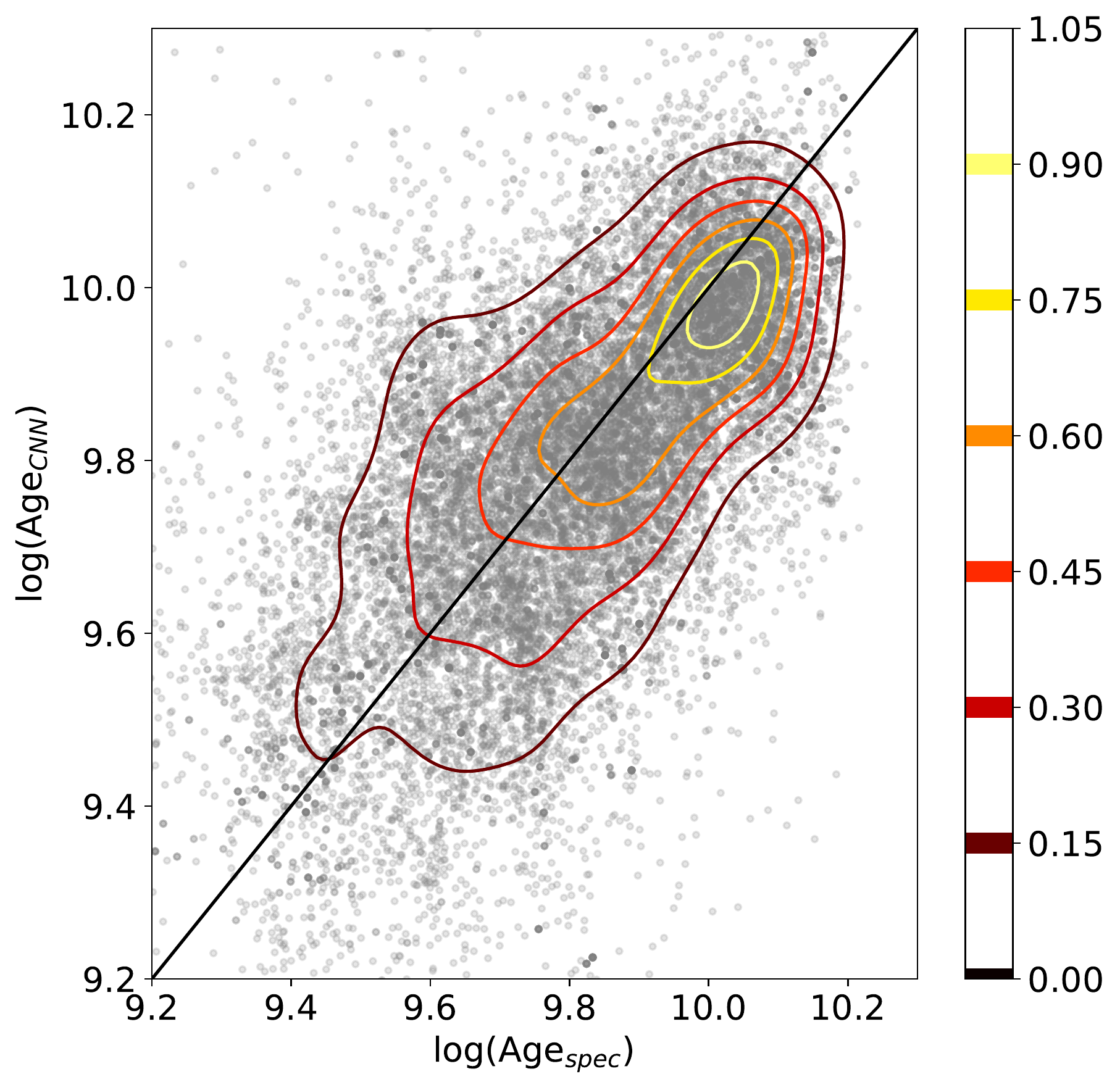}
    \includegraphics[width=\columnwidth]{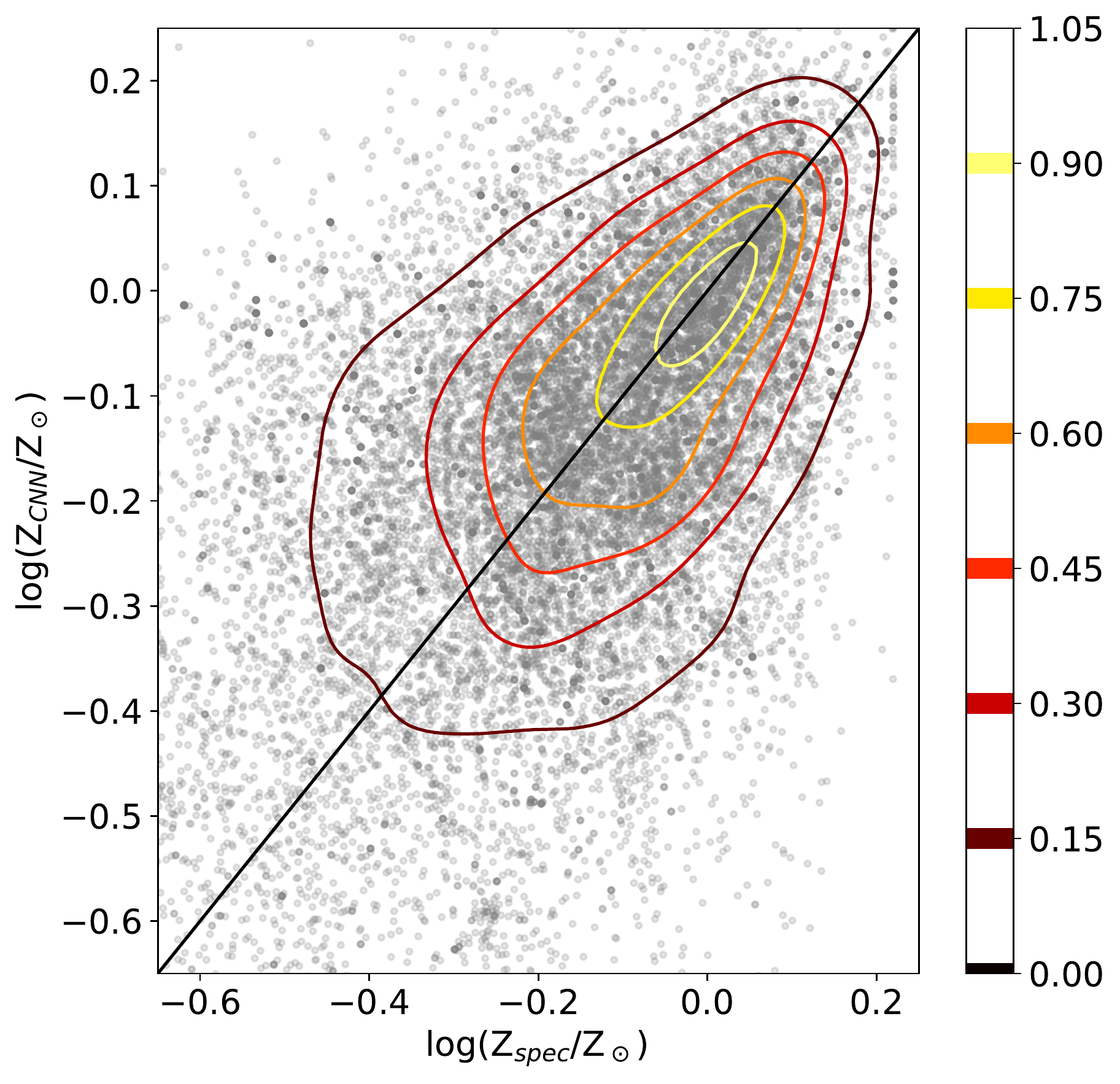}
    \caption{The luminosity-weighted age (Age$_{CNN}$, upper panel) and metallicity (Z$_{CNN}$, lower panel) derived from the CNN against the spectroscopically determined age (Age$_{spec}$) and metallicity (Z$_{spec}$) for Set B. Recovery here is significantly worse than in Set A, with robust standard deviation of 0.16 dex for both age and metallicity.}
    \label{fig:SetBAgeZ}
\end{figure}

The recovery of age and metallicity for Set B is shown in Fig. \ref{fig:SetBAgeZ}. The contour levels are the same as in Fig. \ref{fig:ageZrecovery}. It can be seen that the contours are much more spread out, and not concentrated around the black 1:1 recovery line. The age recovery, in particular, shows an offset with CNN predictions systematically lower than the spectroscopic values. At lower metallicities, the predictions of the CNN become less accurate, which can be seen as the contours spread further from the black 1:1 line. This effect is likely due to the rarity of spectra with $\log( Z_{spec}/Z_\odot) < -0.75$ in the training set. The robust standard deviation in this case are 0.16 dex for both age and metallicity.

\begin{figure}
    \centering
    \includegraphics[width=\columnwidth]{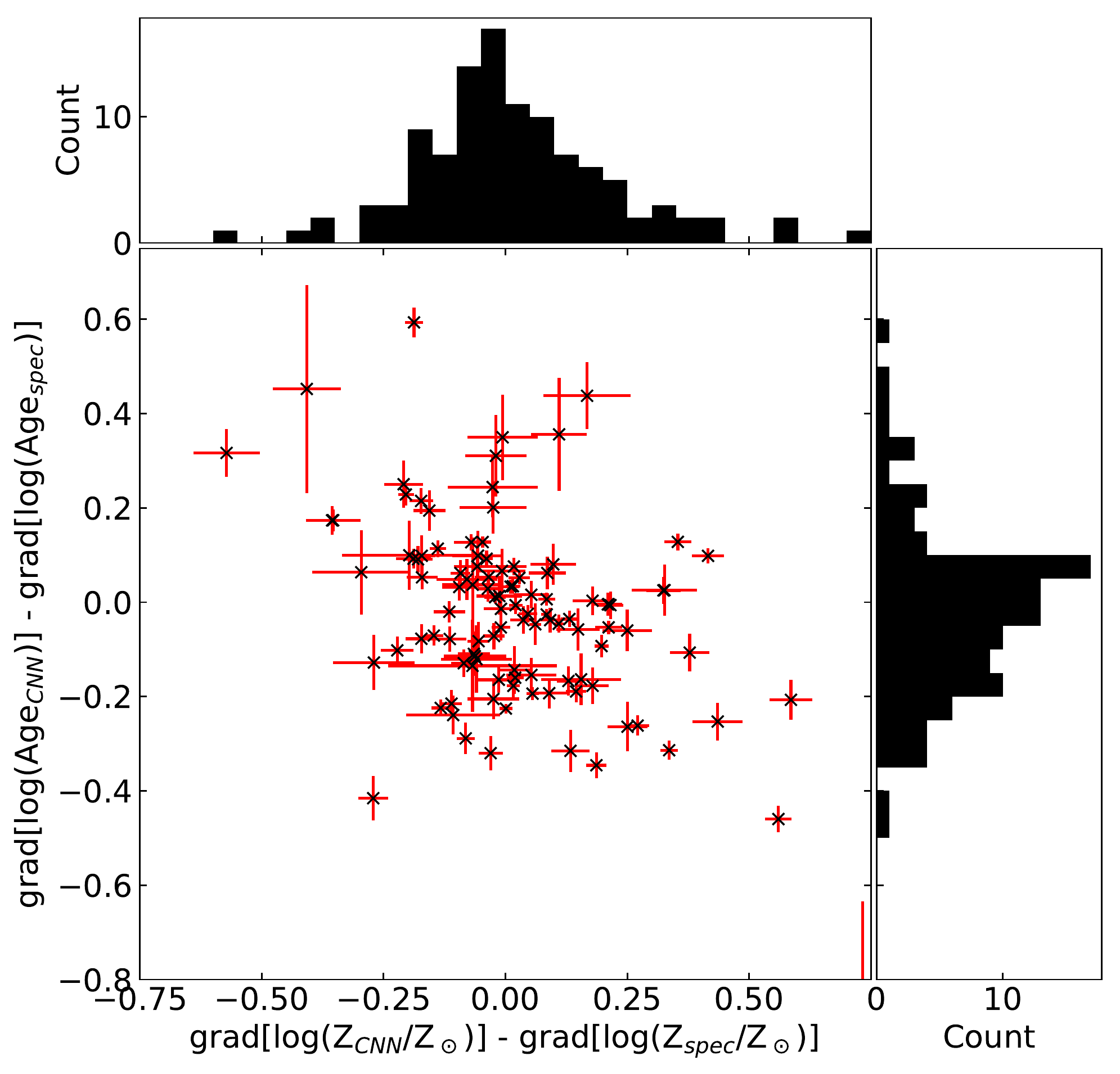}
    \caption{The difference between the gradients from CNN predicted age, grad(log(Age$_{CNN}$)), and the spectroscopically derived age, grad(log(Age$_{spec}$)), against the difference in metallicity gradient from the CNN, grad(log(Z$_{CNN}$/Z$_\odot$)), and spectroscopically derived metallicity, grad(log(Z$_{spec}$/Z$_\odot$)) for Set B. The recovery in Set B is much worse than Set A, with robust standard deviation increased to 0.17 dex/$R_{e}$ and 0.15 dex/$R_{e}$ for age and metallicity, respectively.}
    \label{fig:SetBGradients}
\end{figure}

The quality of the CNN's gradient recovery of the spectroscopic values in Set B are displayed in Fig. \ref{fig:SetBGradients}. These are markedly worse than the results obtained in Set A. In this case, the standard deviation for gradient recovery, grad$_{CNN} - $ grad$_{spec}$, is 0.17 dex/$R_{e}$ and 0.15 dex/$R_{e}$ for age and metallicity, respectively. 
The reason for this discrepancy between Sets A and B is likely due to the diversity in star formation histories among galaxies. The accuracy of Set A implies that the formation history of different regions within the galaxy are similar. As a result, the training set of Set A contains data with similar stellar populations to the testing set, which improved the performance of the CNN. Conversely, the training set for Set B does not contain enough variation to cover the star formation and chemical evolution histories of the unseen galaxies for the CNN to accurately reproduce the spectroscopic values of age and metallicity. 

\subsection{Dependence on Galactic Morphology}
\label{subsec:morph}

To study the importance of the similarity of stellar populations between the training and testing sets, we explore the dependence of the accuracy of CNN predictions of age and metallicity on galaxy morphology. The galaxies in the sample were split by morphology \citep[taken from the SIMABD database, ][]{SIMBAD} giving 44 early-type galaxies and 146 late-type galaxies. CNNs were trained on 33 of the elliptical galaxies and 114 spiral galaxies, respectively. These CNNs were then applied separately to the remaining galaxies in each morphology set. 

The robust standard deviations for the differences between spectroscopic and CNN predicted values are given in Table \ref{tab:Morphology}. It can be seen that predictions for the ages of each of the morphology groups are more accurate when the CNN has been trained on the same morphology group. Additionally, when the CNN has been trained on only early-type galaxies, the age prediction performs best for  early-type galaxies and has a robust standard deviation of 0.12 dex. Prediction of the age and metallicity of late-type galaxies are of similar quality regardless of whether the CNN is trained on early- or late-type galaxies. This is unexpected, but is likely due to the presence of similar stellar populations between early-type galaxies and the bulges of late-type galaxies. Overall, the recovery of early-type galactic properties is significantly better than the full dataset for Set B but is still worse than for Set A. We believe that the increased accuracy in recovery of early-type galaxies is due to the greater degree of similarity between the stellar populations found in early-types than between late-types. This supports our conclusion that the CNN is more capable of predicting age and metallicity values for stellar populations similar to those present in the training set. Therefore, a larger, high-quality dataset would be crucial for future deep learning analysis of stellar populations.

\begin{table}
    \centering
    \caption{The robust standard deviations of the difference between spectroscopic and CNN predicted age (upper) and metallicity (lower table). The columns indicate whether the CNN was trained on early- or late-type galaxies, and the rows indicate whether the application set was composed of early-type or late-type galaxies. See the text for more information.
    }
    \label{tab:Morphology}
    \begin{tabular}{c| l | l l} 
        \hline
        \multirow{2}{*}{\large \textbf{Age}} & & \multicolumn{2}{c}{\textbf{Training Set}}\\
        &  & Early-types & Late-types\\
        \hline
        \textbf{Application} & Early-types & 0.12  & 0.18 \\
        \textbf{Set} & Late-types & 0.20 & 0.19 \\
        \hline
    \end{tabular}
    \begin{tabular}{c| l | l l} 
        \hline
        \multirow{2}{*}{\large \textbf{Z}} & & \multicolumn{2}{c}{\textbf{Training Set}}\\
        &  & Early-types & Late-types\\
        \hline
        \textbf{Application} & Early-types & 0.14 & 0.14 \\
        \textbf{Set} & Late-types & 0.20 & 0.20 \\
        \hline
    \end{tabular}
\end{table}

\subsection{Training set size}
\label{subsec:size}

The size of the training set is very important in neural networks. Typically, very large datasets are used in analysis using CNNs. This is because a large volume of data increases the accuracy of neural network predictions. In this section, we discuss the impact of how the size of the training set affects the predictions of our CNN model, though we are still limited by our relatively small dataset.

\begin{figure}
    \centering
    \includegraphics[width=\columnwidth]{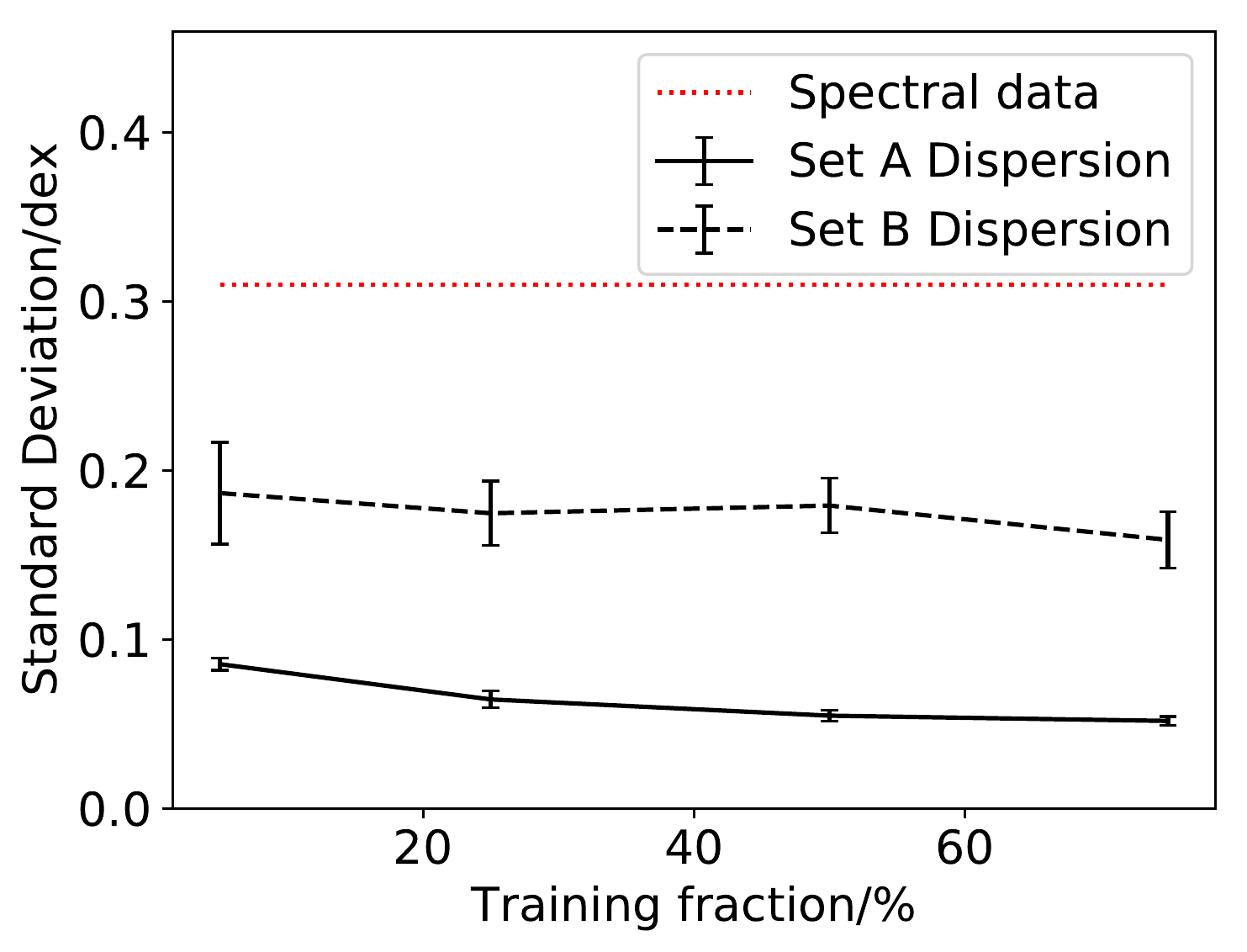}
    \caption{The variation in standard deviation of CNN recovery of age values as a function of the size of the training set is varied. The training set varied between 14795 (75\% of the full dataset) and 986 (5\%) SEDs for Set A (solid line). Training with Set B uses between 157 galaxies  (75\%) and 10 galaxies (5\%) (dashed line). The red dotted line shows the standard deviation we would expect if the predictions were made by simply choosing a random value from the set of spectroscopic ages.}
    \label{fig:RecoverySize}
\end{figure}

Fig. \ref{fig:RecoverySize} shows the robust standard deviation of the difference between spectroscopic and CNN predicted age values for Set A (solid lines) and Set B (dashed lines) as a function of the training set size, given as a fraction of the total size of the dataset. Note that we only used the results for data points whose spectroscopic values are reliable (i.e. with reduced $\chi^2 < 2$), to evaluate the performance when the CNN model is applied to the similar quality data to the training set. 
Training and application of the CNN model was performed 100 times with randomly selected training and application sets for each iteration. The standard deviation for the recovery of age was recorded for each model, and the mean and uncertainty of these standard deviations is shown in Fig \ref{fig:RecoverySize}.
The horizontal red dotted line shows the robust standard deviation of 'predictions' for each spectrum which were randomly selected from the set of spectroscopic ages. Both Set A and Set B results are below this line, which confirms that the CNN learned some relation to map the input features to the output values better than picking a random value from the training set.

It can be seen that the  accuracy of recovery of both Set A and B decreases as the training set size decreases, and the uncertainty of this accuracy increases. Despite this increase, the recovery in Set A with a training set of 5\% of the total dataset is $\sim 0.1$ dex smaller than the recovery of ages
in Set B using 75\% of the dataset. This supports our conclusion that increasing the number of galaxies in our dataset to account for the diversity in star formation histories is crucial in increasing the accuracy of CNN predictions. In other words, the number and diversity of the spectroscopic data used in this paper is not enough for accurate recovery of stellar population parameters from a testing set composed of galaxies that are not included in the training set. We would expect that with data from more galaxies with a diverse range of star formation histories, the accuracy of the recovery for Set B, when using a large training set, would approach that of Set A. 

These findings imply that the stellar populations in different regions within the same galaxy are significantly more similar than stellar populations in different galaxies with the same age and metallicity. Therefore, in order to use CNNs to predict the age and metallicity in a galaxy, we require a very large training dataset, covering the full parameter space of stellar population properties. 

\subsection{Mass dependence of radial gradients}
\label{subsec:MassVGrads}

\begin{figure*}
    \centering
    
\begin{tabular}{c|c|c}
        \includegraphics[width=0.3\linewidth]{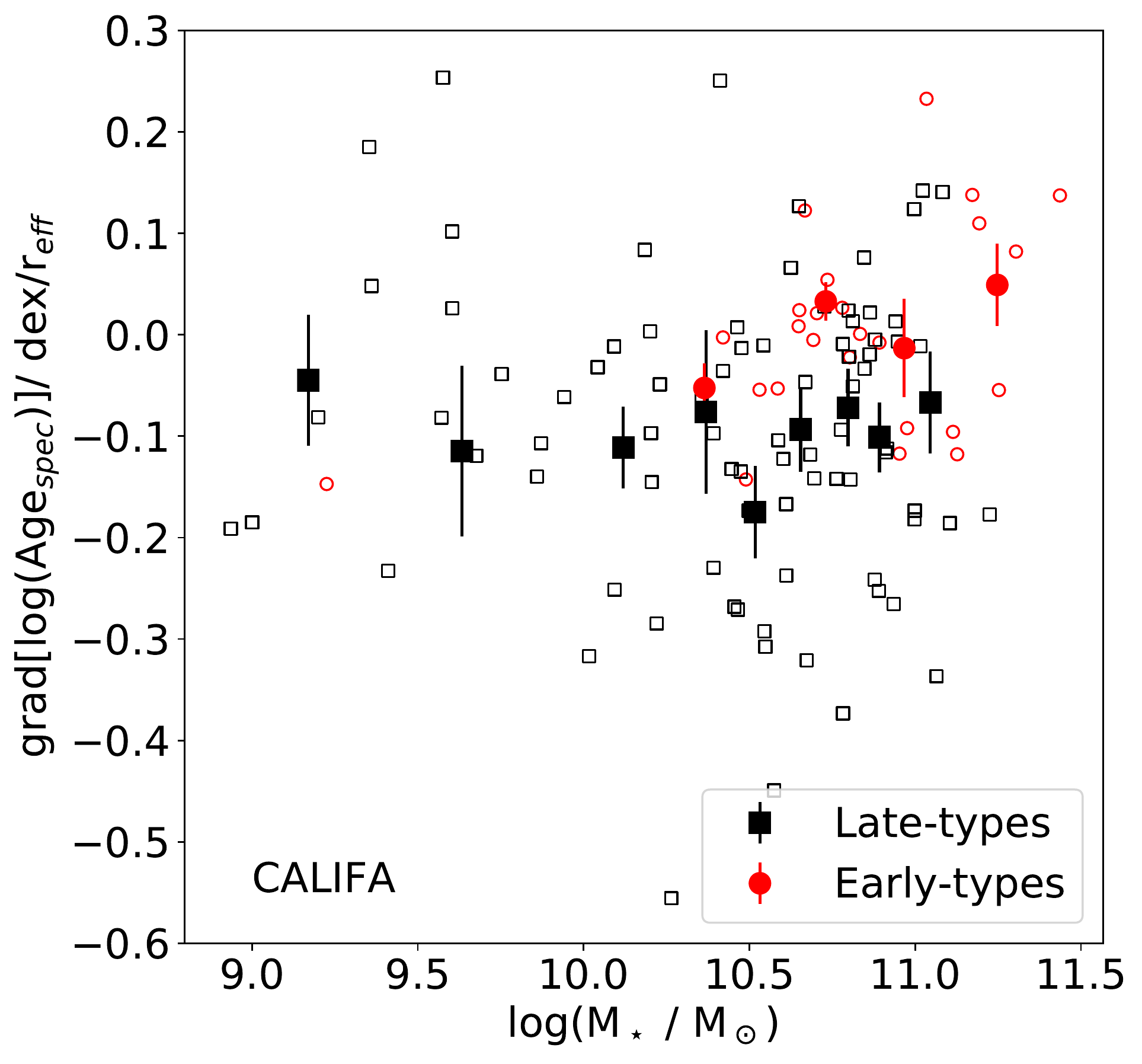} & \includegraphics[width=0.3\linewidth]{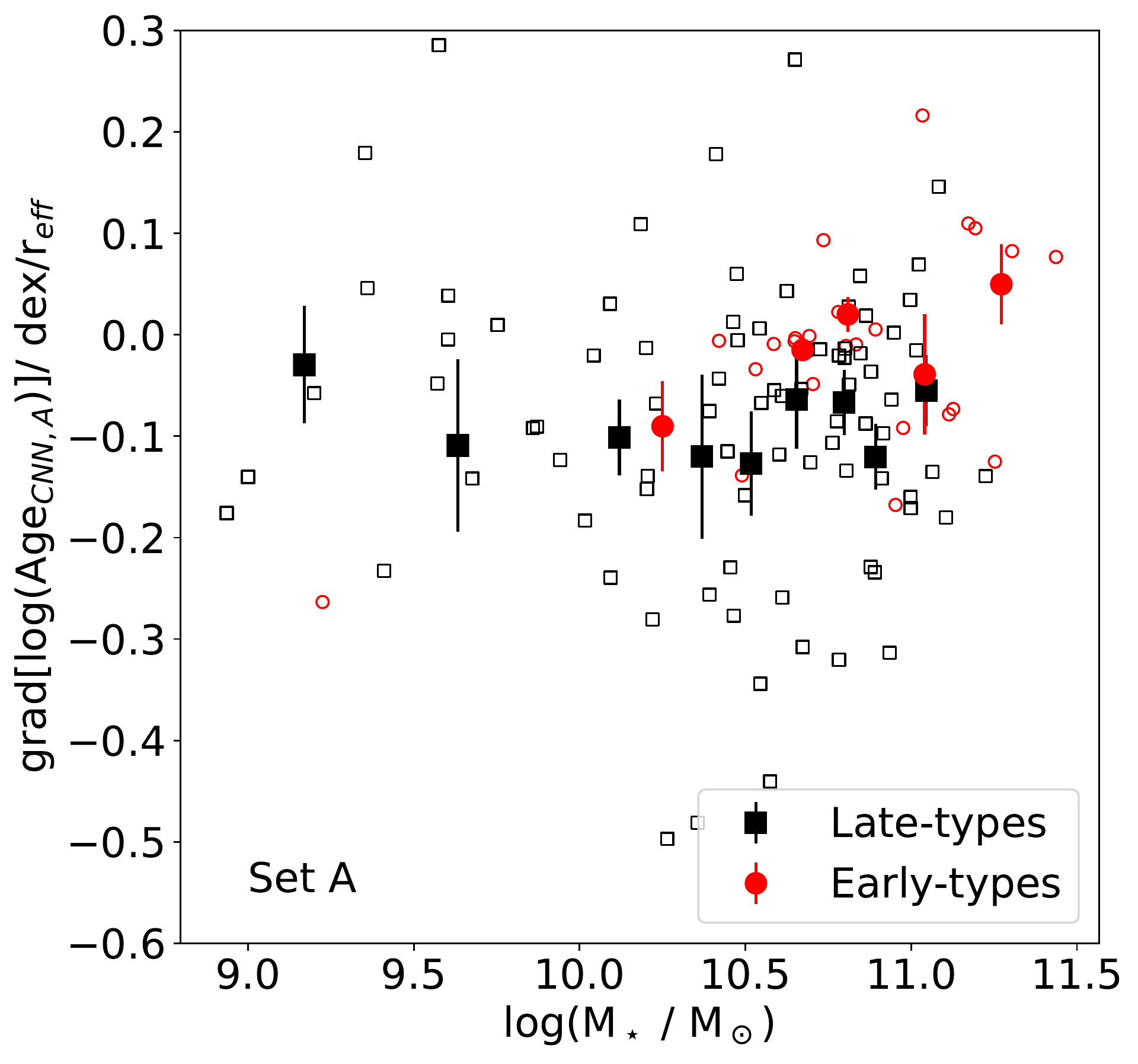} & 
        \includegraphics[width=0.3\linewidth]{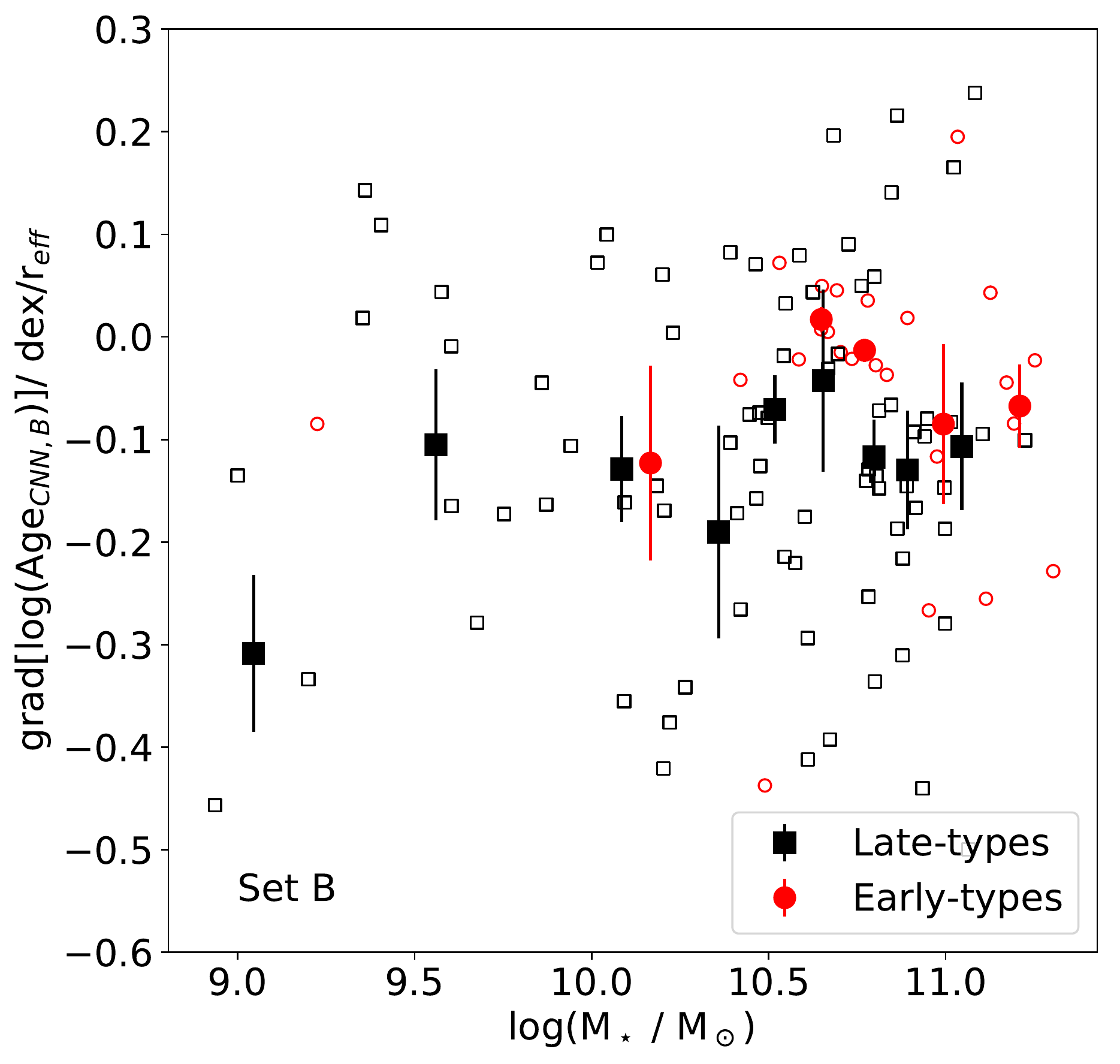}
        \end{tabular}
    
    \caption{The radial age gradient for a galaxy against its stellar mass, using spectroscopically determined gradients from CALIFA (left), and the gradients calculated from CNN predictions with Set A (middle) and Set B (right). The open red circles (open black squares) show the values for individual early- (late-) type galaxies. The filled red circles (filled black squares) show the mean value for each bin of 6 (10) galaxies for early- (late-) type galaxies, with error bars showing the standard deviation. This demonstrates the gradients of Set A are more similar to the spectral gradients than those of Set B.}
    \label{fig:AgeVsMass}
\end{figure*}

    

The dependence of age gradients on galactic stellar mass is of interest when evaluating how galaxies evolve. The relationships we have found between these quantities are shown in Fig. \ref{fig:AgeVsMass}. The left panel in this figure shows the gradients derived from the spectroscopically measured age. The relationship of the late-type (black squares) galaxies' age gradients on mass resembles that of Fig. 6 from \citet{SanchezBlazquez14}, that uses the same spectroscopically derived age values as this paper. This demonstrates that our method of gradient derivation provides consistent results to that of the previous work. It can be seen that the gradients produced by our analysis from Set A (central panel) is 
similar to that of the gradients derived from spectral values (left panel) and therefore showing similar trends to \citet{SanchezBlazquez14}. Conversely, Set B (right panel) shows significant differences from the gradients calculated from 
the spectroscopically derived age (left panel), which can be seen in both the medians for stellar mass bins (filled symbols) and the derived gradient for individual galaxies (open symbols).  

The mass dependence of age gradients for a variety of galactic morphologies was studied in \citet{GonzalezDelgado15}. In Fig. 10 of their paper, the early type galaxies show higher values of the age gradient in the higher mass galaxies at log(M$_{*}$)$\gtrsim 10.5$. The late-type galaxies show similar trends in the same mass range, but show systematically lower gradient than the early-type galaxies. Then, at  log(M$_{*}$)$ \lesssim 10.5$ the gradient values become larger for the smaller mass galaxies in the late-type galaxies. These trends are qualitatively reproduced in the left panel of Fig. \ref{fig:AgeVsMass}. However, the values of the gradients we derived here are systematically higher than those in \citet{GonzalezDelgado15}. This could be due to the differing methods of gradient derivation or differences in stellar population modelling \citep[see][for details]{GonzalezDelgado15}.


\section{Summary and Discussion}
\label{sec:conclusion}

We present a proof of concept study of an application of a CNN model to recover age and metallicity of nearby galaxies. The data used in this work is taken from the CALIFA dataset and is synthesised to produce data resembling 36 J-PAS-like photometric bands which were used to train a CNN model. A total of 21,230 spectra from 190 galaxies are used in this analysis. The CNN was able to predict age and metallicity accurately in Set A (Sections \ref{subsec:reproducing} and \ref{subsec:grads}), where the data used in both the training and application sets came from spectra in the same galaxies. The recovery for age and metallicity is excellent and has a robust standard deviation of 0.05 dex and 0.03 dex, respectively. The radial gradients of age and metallicity are calculated from the CALIFA spectroscopically derived age and metallicity,
and the CNN predictions of these values for each galaxy. The robust standard deviation of the difference between the gradients with spectroscopically derived values and the CNN predicted values
is 0.03 dex/$R_{e}$ for age and 0.02 dex/$R_{e}$ for metallicity. Radial gradients are also recovered well with the CNN. 

On the other hand, for Set B (Section \ref{subsec:trainsets}), where the training and application datasets are composed of spectra located in different galaxies, the CNN's recovery of age and metallicity is markedly worse. The robust standard deviation for the recovery in Set B is a factor of $\sim 3$ worse for age and $\sim 6$ worse for metallicity than Set A. There is also a significant degree of difference between the radial gradients derived from the spectroscopically measured values
and those calculated using predictions from the CNN trained using Set B, due to the greater dispersion of CNN predictions for each spectra. We attribute this decrease in prediction accuracy with respect to Set A to the lesser degree of similarity in stellar populations between different galaxies compared to different regions within the same galaxy. This is supported by the smaller error in recovery for early-type galaxies compared to late-type galaxies in Set B, as the latter have a greater range of stellar populations. Our dataset contains a relatively small number of galaxies, which was not enough to account for the vast diversity of stellar populations. If we had a larger number of galaxies with a great enough overlap of stellar properties, we expect that the CNN predictions would improve greatly and approach the level of accuracy obtained by Set A.

In this work, only the errors from gradient fitting are considered. An improvement to the method would be to consider the error in the CNN predictions of age and metallicity. This would be an important step in properly evaluating the uncertainties of the CNN predictions for the analysis of real observational data. 

We have demonstrated that the CNN model is able to predict age and metallicity values on a relatively small proportion of the training set provided it has enough high quality data to cover the range of stellar populations present in the application set. 
This, along with the low computing power required to apply the trained model to new data, makes CNNs a suitable method of analysis for large datasets such as those that will be produced by the J-PAS survey. 
However, constructing a large enough high-quality training dataset to improve machine learning models is crucial. Therefore, 
we will continue to need additional large spectroscopic surveys and high-performance spectral fitting codes.
More high quality spectral (preferably IFU) data and sophisticated stellar population models to fit these spectra would be invaluable for creating a high quality training set for further neural network studies. 
The efforts in increasing the coverage of IFU surveys, such as SAMI \citep{SAMI} and MaNGA \citep{MaGNA}, and their improving fitting pipelines will be essential in future applications of CNNs to situations similar to that of Set B in this work. 


\section*{Acknowledgements}




\bibliographystyle{mnras}
\bibliography{Paper1} 







\bsp	
\label{lastpage}
\end{document}